\documentclass[11pt]{article}
\usepackage{graphicx}
\usepackage{subfloat}
\DeclareGraphicsExtensions{.pdf}
\usepackage{amsmath}
\usepackage{amsfonts}   
\usepackage{amssymb}

\begin{document}
\date{}
%%%%%%%%%%%%%%%%%%%%
\title{{\bf{\Large Phase transition and scaling behavior of topological charged black holes in Ho\v{r}ava-Lifshitz gravity}}}
%%%%%%%%%%%%%%%%%%%%
\author{
{\bf {{\normalsize Bibhas Ranjan Majhi}}$
$\thanks{e-mail: bibhas@iucaa.ernet.in}}\\
{\normalsize IUCAA,
Post Bag 4, Ganeshkhind,}
\\{\normalsize Pune University Campus, Pune - 411 007, India}\\
and\\
$~${\bf {{\normalsize Dibakar Roychowdhury}}$
$\thanks{e-mail: dibakar@bose.res.in, dibakarphys@gmail.com}}\\
%\\\textit{S.~N.~Bose National Centre for Basic Sciences,}
%\\\textit{JD Block, Sector III, Salt Lake, Kolkata-700098, India}
 {\normalsize S.~N.~Bose National Centre for Basic Sciences,}
\\{\normalsize JD Block, Sector III, Salt Lake, Kolkata-700098, India}
\\[0.3cm]}

%\date{}

\maketitle

%%%%%%%%%%%%%%%%%%%%%%%%%%%%%%%%%%%%%%%%%%%%%%%%%%%%%%%%

\begin{abstract}
  Gravity can be thought as an emergent phenomenon and it has a nice ``thermodynamic'' structure. In this context, it is then possible to study the thermodynamics without knowing the details of the underlying microscopic degrees of freedom. Here, based on the ordinary thermodynamics, we investigate the phase transition of the static, spherically symmetric charged black hole solution with arbitrary scalar curvature $2k$ in Ho\v{r}ava-Lifshitz gravity at the Lifshitz point $z=3$.  The analysis is done using the {\it canonical ensemble} frame work; i.e. the charge is kept fixed. We find (a) for both $k=0$ and $k=1$, there is no phase transition, (b) while $k=-1$ case exhibits the second order phase transition within the {\it physical region} of the black hole. The critical point of second order phase transition is obtained by the divergence of the heat capacity at constant charge. Near the critical point, we find the various critical exponents. It is also observed that they sati
 sfy the usual thermodynamic scaling laws.
\end{abstract}

%%%%%%%%%%%%%%%%%%%%%%%%%%%%%%%%%%%%%%%%%%%%%%%%%%%%%%%%%%%%
\section{Introduction}
   One of the fascinating solutions of the gravity theories is the black hole space time. Pioneering works of Bekenstain \cite{Bekenstein:1973ur}, Hawking \cite{Hawking:1974rv} and later on Bardeen et al \cite{Bardeen} illuminated the fact that all the laws of black hole mechanics are identical to those of the ordinary thermodynamics with properly identification of the respective quantities, like temperature, entropy, energy etc. and hence they do act as the ordinary thermal system.  People then started thinking that black holes may play a crucial role to give idea on the quantum nature of the gravity. But till now there is no such complete theory of quantum gravity. 
   
   Several recent results strongly indicate the possibility that the field equations of gravity have the same status as the equations of fluid mechanics or elasticity. For a recent review, see \cite{paddyaspects}. One specific implementation of this idea considers the field equations of the theory to be `emergent' in a well-defined sense, rather than use that term in a more speculative vein like e.g., considering the space and time themselves to be emergent etc. The evidence for such a specific interpretation comes from different facts like the possibility of interpreting the
field equation in a wide class of theories as thermodynamic relations \cite{thermo}, the nature of action
functional in gravitational theories and their thermodynamic interpretation \cite{holoaction}, the possibility
of obtaining the field equations from a thermodynamic extremum principle \cite{aseemtp}, application of
equipartition ideas to obtain the density of microscopic degrees of freedom \cite{equiv}, the equivalence
of Einstein's field equations to the Navier-Stokes equations near a null surface \cite{NS} etc. The important fact in the emergent paradigm is that one does not need to know about the details of the microscopic description of the theory. These lead to the idea that one can study the different aspects of thermodynamics, in the case of gravity, without any detail of the underlying microscopic structure. Here we will adopt the same logic and study the phase transition of the black hole solution in Ho\v{r}ava-Lifshitz theory following the prescription in ordinary thermodynamics.

  Studying the thermodynamics \cite{hp}-\cite{th17} as well as the critical behavior \cite{cr1}-\cite{cr15} of various black hole solutions in the framework of usual Einstein gravity has been a fascinating topic of research for the past few decades. Apart from these, there also exist topological black hole solutions whose thermodynamics has been studied in Einstein gravity \cite{th18,th19}, Einstein-Gauss-Bonnet gravity and dilaton gravity \cite{th20} as well as Lovelock gravity \cite{th21}. On the other hand, inspired by the dynamical critical phenomena in usual condensed matter systems, very recently
P. Ho\v{r}ava proposed a UV complete theory of gravity \cite{horava} that reduces to the usual Einstein gravity at large scales. Since then a number of attempts have been made in order to understand various aspects of this theory \cite{horava1}-\cite{horava3} including different cosmological aspects \cite{Calcagni:2009ar}. Various black hole solutions were also found in \cite{Lu:2009em,hl3,Tang:2009bu}. The thermodynamics of these black holes have been studied in \cite{hl3,hl1,hl2}.  Although these attempts are self contained and rigorous, still there remains some major questions which have not yet been attempted. Studying the critical behavior of black holes is one of the important aspects, which we aim to explore for  Ho\v{r}ava-Lifshitz theory of gravity. 

In ordinary thermodynamics, critical exponents plays a crucial role in order to understand the singular behavior of various thermodynamic entities near the critical point. In usual thermodynamic systems, it is customary to express the singular behavior of various thermodynamic systems in terms of power laws characterized by a set of \textit{static critical exponents} which determines the qualitative nature of the phase transition near the critical point. All these exponents are not independent and are found to satisfy certain \textit{thermodynamic scaling laws} near the critical point \cite{stanley1}-\cite{stanley2}. It is generally observed that for $ d>4 $, where $ d $ is the spatial dimension of the system, the critical exponents do not depend on the spatial dimension of the system, which reflects the mean field characteristics of a theory. On the other hand if the interaction between the constituent elements are short range then the critical exponents are found to be depe
 ndent on $ d $.
   
   In this paper, based on a canonical framework (i.e; keeping the charge ($ Q $) of the black hole fixed \cite{th3}), we investigate the critical behavior of topological charged black holes in Ho\v{r}ava-Lifshitz theory of gravity at the Lifshitz point $z=3$. The black hole solution is given in \cite{hl3}. Although all the thermodynamic quantities were evaluated earlier \cite{hl3}, a detailed study of the nature of phase transition is still lacking. Particularly the issue regarding the \textit{scaling behavior} of (charged) black holes has never been investigated so far in the frame work of Ho\v{r}ava-Lifshitz theory of gravity. We investigate the phase transition phenomena for all the three cases taking $ k=0,\pm 1 $. We observe the following interesting features:
\vskip 1mm
\noindent
$\bullet$ There is no Hawking Page transition \cite{hp} for black hole with $ k=0,1 $.
\vskip 1mm
\noindent
$\bullet$  For $k=-1$, there is a upper bound in the value of the event horizon, above which the temperature becomes negative. This indicates that above this critical value of horizon, the black hole solution does not exist. We will call this valid range as the {\it physical region}. 
\vskip 1mm
\noindent
$\bullet$ Within the physical region, interesting phase structure could be observed for the hyperbolic charged black holes ($ k=-1 $). In this particular case we observe the second order transition. 
\vskip 1mm
\noindent
This is {\it different} from the usual one. In Einstein gravity, Hawking page transition occurs for $k=1$ while there is no such transition for $k=0,-1$.
The critical point of second order phase transition is marked by the divergence in the heat capacity at constant charge ($ C_Q $). Finally, we explicitly calculate all the \textit{static critical exponents} associated with the second order transition, and check the validity of \textit{thermodynamic scaling laws} near the critical point. Interestingly enough it is found that the hyperbolic charged black holes ($ k = -1 $) in the Ho\v{r}ava-Lifshitz theory of gravity fall under the same universality class to that with the black holes having spherically symmetric topology ($ k = 1 $) in the usual Einstein gravity. The values of these critical exponents indeed suggest a universal \textit{mean field} behavior in black holes which valid in both the Einstein as well as Ho\v{r}ava-Lifshitz theory of gravity \cite{cr14}-\cite{cr15}.

    The plan of the paper is as follows: We begin in section 2 by giving a brief introduction of the black hole solution in Ho\v{r}ava-Lifshitz gravity and use it in section 3 to study the different thermodynamic quantities as well as the phase transition. In section 4, the critical exponents near the critical point are being evaluated and a brief discussion on the validity of the ordinary scaling laws is presented.  Finally, we conclude in section 5.

%%%%%%%%%%%%%%%%%%%%%%%%%%%%%%%%%%%%%%%%%%%%%%%%%%%%%%%%%%%%%
\section{Charged black hole space-time in Ho\v{r}ava-Lifshitz gravity}
     In this section a brief discussion on the black hole solution in Ho\v{r}ava-Lifshitz gravity will be presented. For details, one can follow \cite{hl3} where the meaning of all the parameters are given explicitly. We will mainly concentrate on the solutions at the Lifshitz point $z=3$, particularly, the static spherically symmetric topological charged black holes solution
\begin{equation}
ds^{2}=-f(r)dt^{2}+\frac{dr^{2}}{f(r)}+r^{2}d\Omega_{k}^{2}~,
\end{equation}
where, $d\Omega_{k}^2$ is the line element for a two dimensional Einstein space with constant scalar curvature $2k$. Without loss of generality, one can take $k = 0,\pm 1$ respectively. The form of the metric coefficient, for the detailed balance condition, is given by \cite{hl3}
\begin{equation}
f(r)=k+x^{2}-\sqrt{\alpha x-\frac{q^{2}}{2}}~,
\label{BH}
\end{equation}
where $x = \sqrt{-\Lambda}r$ and $ \Lambda(=-\frac{3}{l^{2}}) $ corresponds to the negative cosmological constant.
The physical mass and the charge ($ Q $) corresponding to the black hole solution are respectively given by, 
\begin{equation}
M=\frac{\kappa^{2}\mu^{2}\Omega_{k}\sqrt{-\Lambda}}{16}\alpha; \,\,\,\ Q=\frac{\kappa^{2}\mu^{2}\Omega_{k}\sqrt{-\Lambda}}{16} q
\end{equation}
where, $ \alpha $, $q$ are the integration constants, $ \Omega_{k} $ is the volume of the two dimensional Einstein space and $ \kappa $, $ \mu $ are the constant parameters of the theory.
%On the other hand, the physical charge ($ Q $) corresponding to the above solution may be expressed as,
%\begin{equation}
%Q=\frac{\kappa^{2}\mu^{2}\Omega_{k}\sqrt{-\Lambda}}{16} q,
%\end{equation}
%where $ q $ is an integration constant.
The event horizon is the solution of the equation $f(r_+) = 0$.

\section{Phase transition}
 In ordinary thermodynamics the phase transition is studied by the divergence of relevant thermodynamic quantities. Here, the same technique will be adopted.  We shall first calculate the Hawking temperature, entropy and specific heat of the black hole using a canonical ensemble frame work, which means that we shall carry out our analysis keeping the total charge ($ Q $) of the black hole fixed \cite{th3}.  Finally, a graphical analysis will be given to study the phase transition. Before we proceed further, let us mention that in the following analysis we re-scale our variables as $ M\rightarrow \frac{M}{\Omega_{k}} $, $ S\rightarrow\frac{S}{\Omega_{k}} $, $ Q\rightarrow \frac{Q}{A \Omega_{k}\sqrt{-\Lambda}} $, where we have set $ 16\pi A=1 $ with $ A= \frac{\kappa^{2}\mu^{2}}{16}$. 

   The Hawking temperature is calculated as
\begin{equation}
T=\frac{f^{'}(r_+)}{4\pi}=\frac{\sqrt{-\Lambda}(3x_{+}^{4}+2kx_{+}^{2}-k^{2}-\frac{Q^{2}}{2})}{8\pi x_{+}(k+x_+^{2})}~,
\label{t}
\end{equation}
and after the above mentioned re-scaling the entropy is found to be
\begin{equation}
S=\int\frac{dM}{T}=\left(\frac{x_{+}^{2}}{4}+\frac{k}{2}lnx_{+} \right)+S_0~.
\label{s}
\end{equation}
In the above, $S_0$ is the integration constant which must be fixed by the physical consideration. To be precise, entropy is always determined upto some additive constant. But in all physical considerations the difference is important. Therefore in the present analysis we will consider only the difference. However, the integration constant $S_0$ could be determined using the usual thermodynamic prescription; i.e. determination of entropy in the $T\rightarrow 0$ limit, which might be the entropy of an extremal black hole.
\begin{figure}[h]
\centering
%\rotatebox{270}{
\includegraphics[angle=0,width=12cm,keepaspectratio]{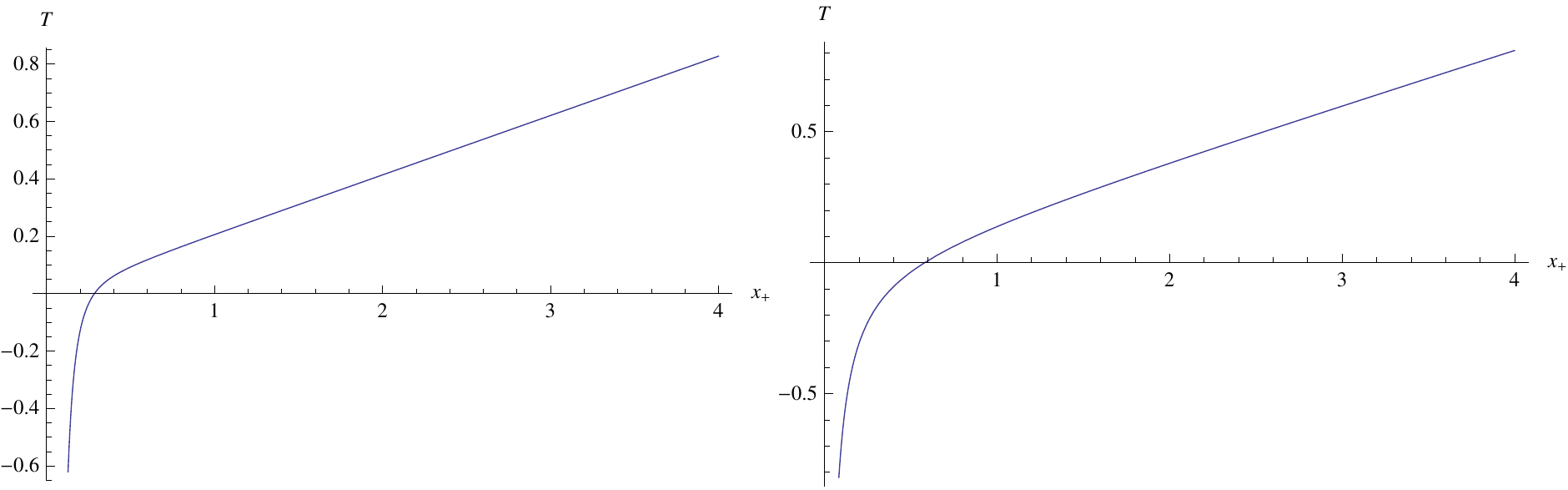}
\caption[]{\it Temperature ($ T $) plot of topological black holes for $ k=0,1 $ with respect to $x_{+}$ for $Q=5$ and $ l=1 $.}
%\label{figure 2b}
\end{figure}

\begin{figure}[h]
\centering
%\rotatebox{270}{
\includegraphics[angle=0,width=12cm,keepaspectratio]{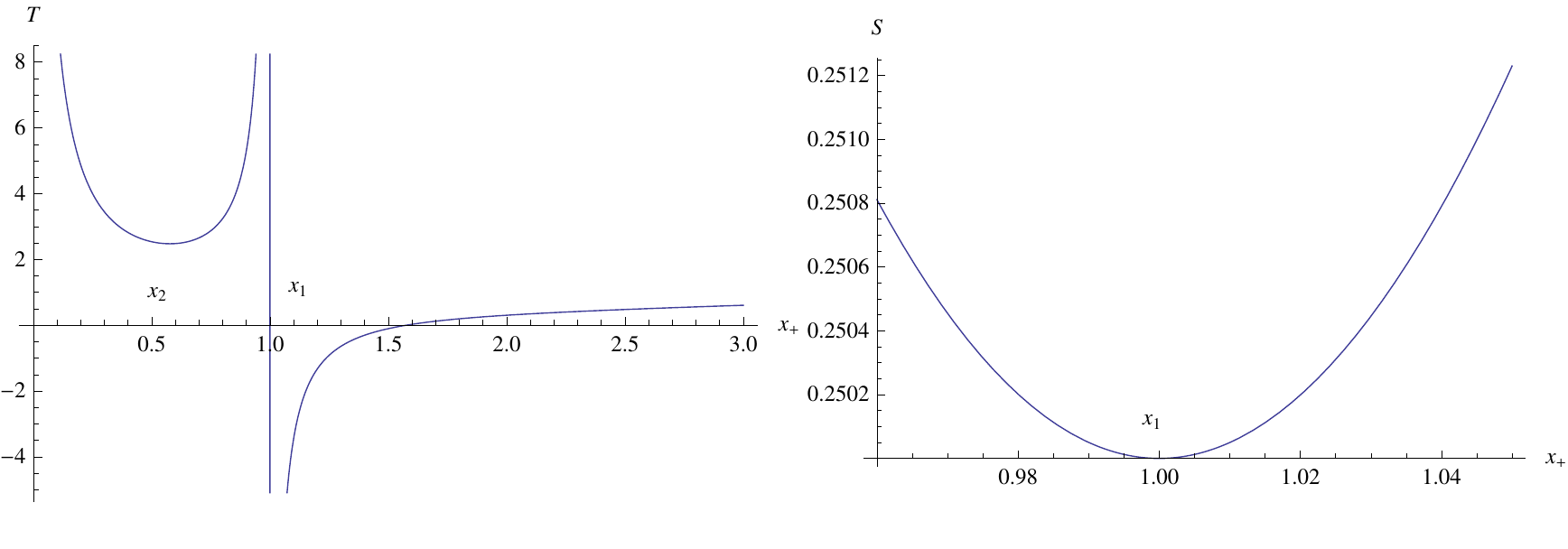}
\caption[]{\it Temperature ($ T $) and Entropy ($ S $) plot of topological black holes for $ k=-1 $ with respect to $x_{+}$ for $Q=5$ and $ l=1 $.}
%\label{figure 2b}
\end{figure}

Before we proceed further, let us first try to analyze the behavior of Hawking temperature ($ T $) for different choices of $ k $. First of all, for the case $ k =0,1 $, we note that the Hawking temperature ($ T $) is a monotonically increasing function of horizon radius ($ x_+ $) (see fig. 1) which indicates that the corresponding heat capacities ($ C_Q $) are always positive definite. Therefore these black holes are globally stable and there is no phase transition.

Next consider the other case, $ k=-1 $. For this, the Hawking temperature (\ref{t}) reduces to the following form: 
\begin{equation}
T = \frac{\sqrt{-\Lambda}(3x_+^{2}+1)}{8\pi x_+} - \frac{\sqrt{-\Lambda}Q^{2}}{16 \pi x_+ (x_+^{2}-1)\label{tk}}.
\end{equation}
It it interesting to note that the Hawking temperature ($ T $) is always positive in the range $ 0<x_+<1 $. Furthermore, if we plot the entropy ($S$) as a function of $x_+$, using (\ref{s}), it shows that $S$ is also positive in this range (see fig. 2). Interestingly enough this condition is found to be valid for any value of the physical charge ($ Q $) of the black hole. On the other hand, we note that $ T\rightarrow -\infty $ as $ x_+\rightarrow 1 $. This is due to the fact that as $ x_+\rightarrow 1 $ the second term on the r.h.s of (\ref{tk}) dominates over the first one which ultimately produces a large negative temperature. This is a nonphysical situation and the corresponding black hole solution does not exist for $ x_+\geq1 $.  Considering this fact, in the present paper we carry out our analysis in the {\it physical range} $ 0<x_+<1 $ where the temperature ($ T $) of the black hole is \textit{finite} as well as \textit{positive} definite. 

   In the above specified range ($ 0<x_+<1 $) we observe a change in slope at $ x_+=x_2 $ of the corresponding ($ T-x_+ $) plot (see fig. 2). This change in slope signals a discontinuity at $ x_+=x_2 $ in the corresponding heat capacity $ C_Q $. It gives a indication of second order phase transition at $ x_+=x_2 $, which we shall refer as the critical point of phase transition. In order to make the discussion more transparent we  next compute the corresponding heat capacity ($ C_Q $).

Using (\ref{t}) and (\ref{s}) the heat capacity is determined as
\begin{eqnarray}
C_Q&=&T\left(\frac{\partial S}{\partial T} \right)_{Q}=T\frac{\left(\frac{\partial S}{\partial x_+} \right)_{Q}}{\left(\frac{\partial T}{\partial x_+} \right)_{Q}}\nonumber\\
&=&\frac{(k+x_{+}^{2})^{2}(3x_{+}^{4}+2kx_+^{2}-k^{2}-\frac{Q^{2}}{2})}{6x_+^{6}+14kx_+^{4}+10k^{2}x_+^{2}+2k^{3}+Q^{2}(k+3x_+^{2})}~.
\label{cq} 
\end{eqnarray}
\begin{figure}[h]
\centering
%\rotatebox{270}{
\includegraphics[angle=0,width=8cm,keepaspectratio]{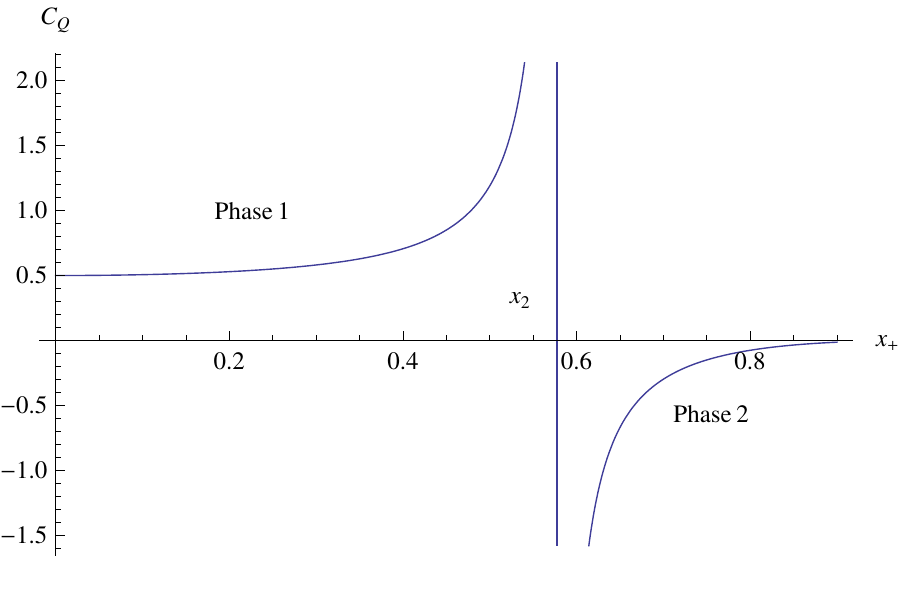}
\caption[]{\it Specific heat ($ C_{Q} $) plot of topological black holes for $ k=-1 $ with respect to $x_{+}$ for $Q=5$ and $ l=1 $.}
%\label{figure 2b}
\end{figure}

From the above figure (fig. 3), we see that specific heat ($ C_Q $) indeed suffers a discontinuity at $ x_+ = x_2 $. This shows that there is a genuine second order phase transition at $ x_+ = x_2 $. For the phase 1 the heat capacity ($ C_Q $) is always found to be positive (see fig. 3) which means that this phase is thermodynamically {\it stable}. On the other hand, $ C_Q<0 $ for the phase 2 and therefore it is an {\it unstable phase}. 

 This scenario is completely different in the Einstein gravity, where we have no Hawking Page transition for $ k=0,-1 $. Whereas, $k=1$ case exhibits Hawking Page transition \cite{th18}-\cite{th20}. For the above phase structure one can further investigate the critical behavior of the black hole in Ho\v{r}ava-Lifshitz theory of gravity near the critical point $x_2$, which we aim to discuss in the next section.

%%%%%%%%%%%%%%%%%%%%%%%%%%%%%%%%%%%%%%%%%%%%%%%%%%%%%%%%%%%%
\section{Critical phenomena and scaling laws}

In this section, based on a thermodynamic approach, we explicitly calculate various critical exponents (that is associated with the second order phase transition at $x_2$) for the hyperbolic charged black holes (with $ k=-1 $) in the fixed charge ($ Q $) ensemble. 

In the theory of phase transitions, critical exponents play an important role in order to understand the singular behavior of various thermodynamic entities near the critical point(s). In order to have a complete understanding of the physics of phase transition phenomena, one generally introduces a set of static critical exponents  $ (\alpha,\beta,\gamma,\delta,\varphi,\psi,\nu,\eta) $ which play a central role in the theory of critical phenomena, and studies the so called thermodynamic scaling laws \cite{stanley1} near the critical point.

   In order to find the critical exponents, we write near the critical point
\begin{eqnarray}
x_+=x_2(1+\Delta),
\label{eq1}
\end{eqnarray}
where $ |\Delta|<<1 $. Defining $T_2 \equiv T(x_2)$,
the Taylor expansion of the temperature ($ T $) about $ x_+=x_2 $ yields
\begin{equation}
T=T_2+\Big[\left( \frac{\partial T}{\partial x_+}\right) _{Q}\Big]_{x_+=x_2}(x_+-x_2)+\frac{1}{2}\Big[\left(\frac{\partial^{2}T}{\partial x_+^{2}} \right)_{Q}\Big]_{x_+=x_2}(x_+-x_2)^{2}+ {\textrm{ higher order terms}}~.
\label{eq2} 
\end{equation}
It has been shown earlier that at the critical point $x_2$, $C_Q$ diverges (see Figure 3). Therefore (\ref{cq}) implies that the second term on the right hand side of (\ref{eq2}) vanishes. Hence neglecting the higher order terms and then using (\ref{eq1}) in (\ref{eq2}) we obtain,
\begin{equation}
\Delta = \sqrt{\frac{2}{D}}\frac{(T-T_2)^{1/2}}{x_2}
\label{eq3}
\end{equation}
where,
\begin{equation}
D=\left[ \left(\frac{\partial^{2}T}{\partial x_+^{2}} \right)_{Q}\right] _{x_+=x_2}=\frac{6x_2^{2}+3Q^{2}x_2^{2}-6x_2^{4}-6Q^{2}x_2^{4}-2-Q^{2}}{x_2^{3}(x_{2}^{2}-1)^{3}}.
\end{equation}

  Now to find the critical exponent $\alpha$ which is defined by the standard relation
\begin{equation}
C_Q\sim |T-T_2|^{-\alpha}~,
\label{alpha}
\end{equation}
we use the relation (\ref{eq1}) in (\ref{cq}) and then keep the terms only linear in $ \Delta $ to obtain
\begin{eqnarray}
C_Q \simeq \frac{\mathcal{N}(x_2,Q)}{\Delta (36x_2^{2}-56x_2^{4}+20x_2^{2}+6Q^{2}x_2^{2})}~,
\label{cq1}
\end{eqnarray}
with
\begin{equation}
\mathcal{N}(x_2,Q)=(x_2^{2}-1)^{2}(3x_2^{4}-2x_2^{2}-1-Q^{2}/2)~.
\end{equation}
Finally, using (\ref{eq3}) in the above we find the behavior of $ C_Q $ near the critical point $ x_+=x_2 $:
\begin{equation}
C_Q\simeq \frac{\mathcal{A}(x_2,Q)}{(T-T_2)^{1/2}}
\label{eq4}
\end{equation}
where,
\begin{equation}
\mathcal{A}(x_2,Q)=\frac{\mathcal{N}(x_2,Q)\sqrt{D}x_2}{\sqrt{2}(36x_2^{2}-56x_2^{4}+20x_2^{2}+6Q^{2}x_2^{2})}~.
\end{equation}
A comparison of (\ref{eq4}) with the standard relation (\ref{alpha}) yields $ \alpha=1/2 $.

  Next the determination of the critical exponent $ \beta $ which is defined through the relation (for a fixed value of charge $ Q $) as,
\begin{equation}
\Phi(x_+)-\Phi(x_2)\sim |T-T_2|^{\beta}
\label{eq5}
\end{equation}
will be done. Here the potential $\Phi(x)$ is given by
\begin{eqnarray}
\Phi(x) = \frac{Q}{x}~.
\label{potential}
\end{eqnarray}
To proceed, let us first Taylor expand $ \Phi(x_+)=\frac{Q}{x_+}$ about $x_+ = x_2$:
\begin{eqnarray}
\Phi(x_+) &=& \Phi(x_2) + \Big[\Big(\frac{\partial\Phi}{\partial x_+}\Big)_{Q}\Big]_{x_+=x_2} (x_+-x_2)+{\textrm{higher order terms}}
\nonumber
\\
&=&  \Phi(x_2)  - \frac{Q}{x_2^2} (x_+-x_2)+{\textrm{higher order terms}}~.
\label{phi1}
\end{eqnarray}
As before, neglecting the higher order terms and then using (\ref{eq1}) we obtain,
\begin{equation}
\Phi(x_+)-\Phi(x_2)=-\frac{Q}{x_2^{2}}\sqrt{\frac{2}{D}}(T-T_2)^{1/2}~,
\label{eq6}
\end{equation}
where in the final step (\ref{eq3}) also has been used.
This immediately determines $ \beta=1/2 $.

   To find out the exponent $ \gamma $, associated with the divergence of the inverse of the isothermal compressibility $ K_T^{-1} = \Big[Q\left(\frac{\partial \Phi}{\partial Q} \right)_{T}\Big]$ \cite{cr6}, we first find the explicit expression of $K_T^{-1}$. The thermodynamic identity $\left(\frac{\partial \Phi}{\partial T} \right)_{Q} \left(\frac{\partial T}{\partial Q} \right)_{\Phi} \left(\frac{\partial Q}{\partial \Phi} \right)_{T}=-1$ yields
\begin{eqnarray}
\Big(\frac{\partial\Phi}{\partial Q}\Big)_{T} = -\Big(\frac{\partial\Phi}{\partial T}\Big)_{Q}\Big(\frac{\partial T}{\partial Q}\Big)_{\Phi}.
\label{KT1}
\end{eqnarray}
In order to evaluate the right hand side of (\ref{KT1}) first note that,
\begin{equation}
\Big(\frac{\partial\Phi}{\partial T}\Big)_{Q} = \frac{\Big(\frac{\partial\Phi}{\partial x_+}\Big)_{Q}}{\Big(\frac{\partial T}{\partial x_+}\Big)_{Q}}.\label{neweq1}
\end{equation}
Also, from the functional relation,
\begin{equation}
T=T(x_+,Q)
\end{equation}
we find,
\begin{equation}
\Big(\frac{\partial T}{\partial Q}\Big)_{\Phi} = \Big(\frac{\partial T}{\partial x_+}\Big)_{Q} \Big(\frac{\partial x_+}{\partial Q}\Big)_{\Phi} + \Big(\frac{\partial T}{\partial Q}\Big)_{x_+}.\label{neweq2}
\end{equation}
Right hand side of both (\ref{neweq1}) and (\ref{neweq2}) can be easily calculated using the relations (\ref{t}) and (\ref{potential}) which finally yields,
\begin{equation}
K_T^{-1}=Q\left(\frac{\partial \Phi}{\partial Q} \right)_{T} = \left( \frac{Q}{x_+}\right) \frac{6x_{+}^{6}+10x_+^{2}-14x_+^{4}-2+Q^{2}(x_+^{2}+1)}{6x_+^{6}+10x_+^{2}-14x_+^{4}-2+Q^{2}(3x_+^{2}-1)}.
\end{equation}
Then to obtain the near critical point expression, use (\ref{eq1}) and next (\ref{eq3}) in the above. This yields,
\begin{equation}
K_T^{-1}\simeq \frac{Q\sqrt{D}(6x_2^{6}+10x_2^{2}-14x_2^{4}-2+Q^{2}(x_2^{2}+1))}{\sqrt{2}(36x_2^{6}-56x_2^{4}+20x_2^{2}+6Q^{2}x_2^{2})}(T-T_2)^{-1/2}.
\end{equation}
Comparing with the standard relation 
$K_T^{-1}\sim |T-T_2|^{-\gamma}$,
defined for the fixed value of charge ($ Q $), we get $ \gamma =1/2 $.

   The critical exponent $ \delta $ is defined through the relation,
\begin{equation}
\Phi(x_+)-\Phi(x_2)\sim |Q-Q_2|^{1/\delta}
\label{delta}
\end{equation}
at constant temperature $ T $.
To find it, we first expand $ Q(x_+) $ in a sufficiently small neighborhood of $x_+= x_2 $ which yields,
\begin{eqnarray}
Q(x_+)&=& Q(x_2)+\left[ \left( \frac{\partial Q}{\partial x_+}\right)_{T}\right]_{x_+=x_2} (x_+-x_2)+\frac{1}{2} \left[ \left( \frac{\partial^{2} Q}{\partial x^{2}_+}\right)_{T}\right]_{x_+=x_2} (x_+-x_2)^{2}
\nonumber
\\
&+& {\textrm{higher order terms}}~.
\label{Q}
\end{eqnarray}
Since $ T=T(x_+,Q) $ (see Eq. (\ref{t})), one finds for the fixed $T$
\begin{equation}
\left( \frac{\partial Q}{\partial x_+}\right)_{T}=- \left( \frac{\partial T}{\partial x_+}\right)_{Q} \left( \frac{\partial Q}{\partial T}\right)_{x_+}~.
\label{eq7}
\end{equation}
Now at the critical point $x_+ = x_2$, $C_Q$ diverges and so, as earlier, $\Big[\left( \frac{\partial T}{\partial x_+}\right)_{Q}\Big]_{x_+=x_2}=0 $. Therefore, after neglecting the higher order terms, (\ref{Q}) reduces to
\begin{equation}
x_+ -x_2=\sqrt{\frac{2}{M}}(Q(x_+)-Q(x_2))^{1/2}\label{eq8}
\end{equation}
where,
\begin{equation}
M=\left[ \left( \frac{\partial^{2} Q}{\partial x^{2}_+}\right)_{T}\right]_{x_+=x_2}=\frac{18x_2^{8}+32x_2^{4}-44x_2^{6}-3Q^{2}x_2^{4}-4x_2^{2}-2-Q^{2}}{2Q^{2}x_2^{2}(x_2^{2}-1)^{2}}.
\end{equation}
Next since $\Phi(x_+) = \frac{Q}{x_+}$, we have the functional relation
$\Phi=\Phi(x_+,Q)$ and so
\begin{eqnarray}
\Big(\frac{\partial \Phi}{\partial x_+}\Big)_T = \Big(\frac{\partial\Phi}{\partial x_+}\Big)_Q + \Big(\frac{\partial\Phi}{\partial Q}\Big)_{x_+} \Big(\frac{\partial Q}{\partial x_+}\Big)_T~.
\label{phi2}
\end{eqnarray}
Therefore using (\ref{eq7}) we find,
\begin{equation}
\left[ \left( \frac{\partial \Phi}{\partial x_+}\right)_{T} \right]_{x_+=x_2} = \left[ \left( \frac{\partial \Phi}{\partial x_+}\right)_{Q} \right]_{x_+=x_2}=-\frac{Q}{x_2^{2}}.\label{eq9}
\end{equation}  
Finally, expanding $ \Phi(x_+) $ close to the critical point $ x_+\sim x_2 $ at constant $T$ and then using (\ref{eq8}) and (\ref{eq9}) we obtain,
\begin{equation}
\Phi(x_+)-\Phi(x_2)\simeq -\frac{Q}{x_2^{2}} \sqrt{\frac{2}{M}}\Big[Q(x_+)-Q(x_2)\Big]^{1/2}~.
\label{eq10}
\end{equation} 
Hence, the critical exponent is read off as $ \delta =2 $.

   Now re-expressing (\ref{eq4}) by using (\ref{eq3}) in the following form:
\begin{equation}
C_Q = \sqrt{\frac{2}{D}}\frac{{\cal{A}}}{x_2\Delta} =  \sqrt{\frac{2}{D}}\frac{{\cal{A}}}{(x_+-x_2)}~,
\end{equation}
and then substituting the value of $x_+-x_2$ from (\ref{eq8}) we obtain\footnote{$ Q(x_2)=Q_2 $.}
\begin{equation}
C_Q\sim \frac{1}{(Q(x_+)-Q_2)^{1/2}}~.
\label{eq11}
\end{equation}
Comparing this with the standard definition
$C_Q \sim {|Q(x_+)-Q_2|^{-\varphi}}$
we find $ \varphi=1/2 $.

 In the following the calculation, the critical exponent $ \psi $, defined by
\begin{equation}
S(x_+)-S(x_2)\sim |Q-Q_2|^{\psi}~,
\end{equation}
will be found out. Expansion of the entropy for fixed charge $Q$ about the critical point $x_2$ yields
\begin{eqnarray}
S(x_+) = S(x_2) + \Big[\Big(\frac{\partial S}{\partial x_+}\Big)_Q\Big]_{x_+ = x_2} (x_+ - x_2)+{\textrm{higher order terms}}~.
\label{new1}
\end{eqnarray}
Then following the identical steps as earlier and using (\ref{eq8}) we find
\begin{equation}
S(x_+)-S(x_2)\sim \frac{(x_2^{2}-1)}{2x_2}\sqrt{\frac{2}{M}}(Q-Q_2)^{1/2}~,
\end{equation}
which yields $ \psi = 1/2 $.

 In the following table, we give the values of the critical exponents for the present example:
\begin{table}[h]
\caption{Various critical exponents and their values}   
\centering                          
\begin{tabular}{c c c c c c c c c}            
\hline\hline                        
$Critical~Exponents  $ & $\alpha$ & $\beta$  & $\gamma$ & $ \delta $ & $\varphi $ & $ \psi $ \\ [2.0ex]
\hline
Values & 1/2 & 1/2 & 1/2 & 2 & 1/2 & 1/2\\ [2.0ex]         
\hline                              
\end{tabular}
\label{E1}          
\end{table}
It may be interesting to mention that the critical exponents  for the second order phase transition in the case of hyperbolic black hole in Ho\v{r}ava-Lifshitz gravity are exactly equal to those obtained earlier in \cite{cr14} for the Einstein-Born-Infeld gravity with $k=1$.

In the last part of the section, we discuss about the validity of the usual thermodynamic scaling laws for our present situation. It should be mentioned that in the realistic thermodynamic systems various critical exponents are found to satisfy certain (scaling) relations among themselves, known as \textit{thermodynamic scaling laws} \cite{stanley1}-\cite{stanley2}, which may be expressed as,
\begin{eqnarray}
\alpha + 2\beta +\gamma = 2,~~~\alpha +\beta(\delta + 1) =2,~~~(2-\alpha)(\delta \psi -1)+1 =(1-\alpha)\delta ~~\nonumber\\
~~ \gamma(\delta +1) =(2-\alpha)(\delta -1),~~~\gamma =\beta (\delta - 1),~~~\varphi + 2\psi -\delta^{-1} =1 \label{scaling laws}.
\end{eqnarray} 
Since the critical exponents, we find here, are exactly equal to those obtained in the usual Einstein-Born-Infeld gravity for $ k=1 $ \cite{cr14}, 
all the above relations (\ref{scaling laws}) are indeed satisfied for the present case, in spite of difference in the nature of the black hole solution and the phase structure. This can be checked easily by substituting the values from the table in (\ref{scaling laws}). In this sense, it may be possible that the black holes in both type of theories fall under the same class such that their critical exponents are identical. 

\noindent
Finally we are in a position to check the additional scaling laws for the exponents $ \nu $ and $ \eta $ which are associated with the diverging nature of correlation length and correlation function near the critical point. Since the spatial dimension ($ d $) of our theory is $ \leq 4 $, therefore it will be natural to assume that the additional scaling relations,
\begin{equation}
\gamma = \nu (2-\eta),~~~ 2-\alpha = \nu d \label{add}
\end{equation}
will hold in general. This immediately determines the other exponents. Taking $ d=3 $ and using the exponent values from table 1, we finally obtain,
\begin{equation}
\nu =1/2,~~~ \eta =1.
\end{equation}
In practical situation, the assumption that the scaling laws (\ref{add}) are valid, may not be true. Hence it must be checked that these are indeed valid or not. Alternatively, one should find the above exponents by some alternative method. 
%%%%%%%%%%%%%%%%%%%%%%%%%%%%%%%%%%%%%%%%%%%%%%%%%%%%%%%%%%%%%%%

\section{Conclusions}
  It is now evident that gravity and thermodynamics are closely connected to each other \cite{Bekenstein:1973ur,Hawking:1974rv,Bardeen}. The repeated failure to quantize gravity led to a parallel development where gravity is believed to be an emergent phenomenon just like thermodynamics and hydrodynamics instead treating it as a fundamental force \cite{paddyaspects}. The fundamental role of gravity is replaced by thermodynamic interpretations leading to similar or equivalent results without knowing the underlying microscopic details.

 In this paper, we adopted the standard thermodynamic approach to explore the phase structure of the topological charged black holes in Ho\v{r}ava-Lifshitz gravity for the Lifshitz point $z=3$. These black holes were introduced earlier  in \cite{hl3}. In \cite{hl3} the authors have computed the general expression for the Hawking temperature for these black holes. In spite of this  attempt, till date the issue regarding the nature of phase transition, particularly the issue of critical phenomena, remains completely unexplored. In the present paper, we attempted to provide an answer to all these questions. A number of interesting features have been observed in this regard which have never been explored so far to the best of our knowledge. These are as follows: 
\vskip 2mm
\noindent
$\bullet$  We showed that there exits a {\it physical range} $0<x_+<1$, where the black hole solution exists.
\vskip 2mm
\noindent 
$\bullet$  Within this range, the second order phase transition occurs.
\vskip 2mm
\noindent
$\bullet$ The situation is not exactly identical to Einstein theory. In Einstein theory, there is no Hawking Page phase transition for $k=0,-1$ while $k=1$ case exhibits such phase transition.  Here, only $k=-1$ solution has phase transition.
\vskip 2mm
\noindent
$\bullet$ Finally, the critical exponents near the critical point of the phase transition were derived. This was done again using the ordinary thermodynamic analogy. It may be noted that the critical point has been marked by the discontinuity of the heat capacity ($C_{Q}$) which suggests that it is a second order phase transition. Moreover, as a point of conformation, we also found that the critical exponent, associated with the divergence in $C_Q$, is $1/2$ which simultaneously indicates a mean field feature as well as  a second order nature of the phase transition. Interestingly, these are exactly equal to those of Einstein-Born-Infeld theory \cite{cr14}-\cite{cr15} and also they satisfy the ordinary thermodynamic scaling laws. This in fact suggests two remarkable facts: (1) black holes in both the Einstein as well as Ho\v{r}ava-Lifshitz theory of gravity fall under the same universality class, (2) there exists a universal \textit{mean field} behavior in both of these gravi
 ty theories. 
   
  It may be pointed out that the issues, dealt with the present paper, are notoriously difficult to study. So far what we have done is completely based on the ordinary thermodynamic analogy. But a simple analogy with the thermodynamics may not be enough to draw safe conclusions. Also the understanding of the thermodynamic interpretation of gravity is far from complete since the arguments are more heuristic than concrete and depend upon specific ansatz or assumptions. Therefore the results are suggestive rather than definitive. Hence it is necessary to compute the critical exponents by other alternative/direct procedures; e.g. correlation function technique of the scalar modes, the  AdS/CFT correspondence, re-normalization group scheme, etc. which is beyond the scope of the present paper. In spite of this, we still believe that the analysis, presented here, could illuminate the underlying microscopic structure of the gravity.

\vskip 5mm

{\bf{ Acknowledgement:}}\\
 D.R would like to thank the Council of Scientific and Industrial Research (C. S. I. R), Government of India, for financial help.

%%%%%%%%%%%%%%%%%%%%%%%%%%%%%%%%%%%%%%%%%%%%%%%%%%%%%%%%%%%%%%%%
  
%%%%%%%%%%%%%%%%%%%%%%%%%%%%%%%%%%%%%%%%%%%%%%%%%%%%%%%%%%%%%%%%%%%%%%%


\begin{thebibliography}{99}
\bibitem{Bekenstein:1973ur}
  J.~D.~Bekenstein,
  %``Black holes and entropy,''
  Phys.\ Rev.\  D {\bf 7}, 2333 (1973).
  %%CITATION = PHRVA,D7,2333;%%
\bibitem{Hawking:1974rv}
  S.~W.~Hawking,
  %``Black hole explosions,''
  Nature {\bf 248}, 30 (1974).
  %%CITATION = NATUA,248,30;%%
\bibitem{Bardeen}
  J.~M.~Bardeen, B.~Carter and S.~W.~Hawking,
  Commun.\ Math.\ Phys.\  {\bf 31}, 161 (1973).
\bibitem{paddyaspects}
T.~Padmanabhan (2010), 
 \emph{Rept. Prog. Phys.}, \textbf{73}, 046901 [arXiv:0911.5004].\\
T. Padmanabhan, \textit{ J.Phys. Conf.Ser.}, \textbf{306},  012001 (2011) [arXiv:1012.4476]. 
\bibitem{thermo} 
T. Padmanabhan,  \textit{Class.Quan.Grav.}, \textbf{19}, 5387 (2002) [gr-qc/0204019];\\ 
Dawood Kothawala and   T. Padmanabhan,  \textit{Phys. Rev.},  \textbf{ D 79} , 104020 (2009) [arXiv:0904.0215];\\
T. Padmanabhan, \textit{A Physical Interpretation of Gravitational Field Equations}\textit{AIP Conference Proceedings},  \textbf{1241},  93-108 (2010) [arXiv:0911.1403].
\bibitem{holoaction} 
 T. Padmanabhan,  \textit{Brazilian Jour.Phys. (Special Issue)}  \textbf{35}, 362 (2005) [gr-qc/0412068];\\
A. Mukhopadhyay and  T. Padmanabhan,  \textit{Phys. Rev.}, \textbf{D 74}, 124023 (2006) [hep-th/0608120];\\
Sanved Kolekar and T. Padmanabhan, \emph{Phys. Rev.},\textbf{ D82}, 024036, 2010 [arXiv:1005.0619].
\bibitem{aseemtp} 
T. Padmanabhan and Aseem Paranjape,  \textit{ Phys.Rev.}  \textbf{ D75} 064004, (2007) [gr-qc/0701003]; \\
T. Padmanabhan,  \textit{ Gen.Rel.Grav}.,  \textbf{ 40}, 529-564 (2008) [arXiv:0705.2533]. 
\bibitem{equiv} 
T. Padmanabhan,  \textit{Mod.Phys.Letts.}, \textbf{ A 25}, 1129-1136 (2010) [arXiv:0912.3165]; \\
T. Padmanabhan,   \textit{ Phys.Rev.}, \textbf{D 81}, 124040 (2010) [arXiv:1003.5665]. 
\bibitem{NS} T. Padmanabhan,   Phys.Rev., \textbf{D 83}, 044048 (2011)  [arXiv:1012.0119]; \\
Sanved Kolekar, T. Padmanabhan, \textit{Action principle for the Fluid-Gravity correspondence and emergent gravity,}
[arXiv:1109.5353] 
%\cite{Iyer:1994ys}
\bibitem{hp}S.W. Hawking, D.N. Page, Comm. Math. Phys. 87, 577 (1983). 
\bibitem{th1} Y. S. Myung, Phys. Lett. B 624 (2005) 297–303.
\bibitem{th2} S. Nojiri, S. D. Odintsov, Phys. Lett. B 521 (2001) 87–95.
\bibitem{th3}A. Chamblin, R. Emparan, C. V. Johnson, R. C. Myers, Phys. Rev. D, 60, 064018, (1999).
\bibitem{th4} S. Fernando, Phys. Rev. D 74, 104032 (2006).
\bibitem{th5}R. A. Konoplya, A. Zhidenko, Phys. Rev. D 78, 104017 (2008).
\bibitem{th6}Y. S. Myung, Y.W. Kim, Y.J. Park, Phys. Rev. D 78, 084002 (2008).
\bibitem{th7} Yun Soo Myung, Phys. Lett. B 638 (2006) 515–518.
\bibitem{th8}Y. S. Myung, Mod.Phys.Lett.A 23,667-676,2008.
\bibitem{th9}B. M. N. Carter, I. P. Neupane, Phys. Rev. D 72, 043534 (2005).
\bibitem{th10} R. G. Cai, Phys. Rev. D 65, 084014 (2002).
\bibitem{th11} R. G. Cai, D. W. Pang, A. Wang, Phys. Rev. D 70, 124034 (2004).
\bibitem{th12} R.~Banerjee, S.~K.~Modak, S.~Samanta, Eur.\ Phys.\ J.\  C  70, 317 (2010).
\bibitem{th13} R.~Banerjee, S.~Ghosh, D.~Roychowdhury, Phys.Lett.B  696, 156 (2011).
\bibitem{th14} R. Banerjee, S. K. Modak and S. Samanta, Phys.Rev.D 84, 064024 (2011) .
\bibitem{th15} R. Banerjee, S. K. Modak and D. Roychowdhury, arXiv:1106.3877 [gr-qc].
\bibitem{th16} R. Banerjee,  D. Roychowdhury, JHEP 11, 004 (2011).
\bibitem{th17} A. Lala, D. Roychowdhury, arXiv:1111.5991 [gr-qc] .
\bibitem{cr1} C.O. Lousto, Nucl. Phys. B 410, 155-172 (1993), gr-qc/9306014.
\bibitem{cr2} C.O. Lousto, Gen. Relativ. Gravit. 27, 121 (1995).
\bibitem{cr3} R. G. Cai, Y.S. Myung, Nucl. Phys. B 495, 339-362 (1997).
\bibitem{cr5} C.O. Lousto, Int.J.Mod.Phys. D 6, 575-590 (1997).
\bibitem{cr6} C.O. Lousto, Phys. Rev. D 51, 1733 (1995), gr-qc/9405048.
\bibitem{cr7} K. Maeda, M. Natsuume, T. Okamura, Phys. Rev. D 78, 106007 (2008).
\bibitem{cr8}  S. Jain, S. Mukherji, S. Mukhopadhyay, JHEP 0911:051 (2009), arXiv:0906.5134 [hep-th]. 
\bibitem{cr9} A. Sahay, T. Sarkar, G. Sengupta, JHEP 1007:082,(2010), arXiv:1004.1625 [hep-th]. 
\bibitem{cr10} A. Sahay, T. Sarkar, G. Sengupta, JHEP 1011:125, (2010), arXiv:1009.2236 [hep-th] .
\bibitem{cr11} Y. Liu, Q. Pan, B. Wang, R. G. Cai, Phys. Lett. B 693, 343–350 (2010), arXiv:1007.2536 [hep-th].
\bibitem{cr12}S. Carlip, S. Vaidya,  Class. Quantum Grav. 20 , 3827-3837 (2003), gr-qc/0306054. 
\bibitem{cr13} X. N. Wu, Phys. Rev. D 62, 124023 (2000).
\bibitem{cr14} R. Banerjee, D. Roychowdhury, Phys. Rev. D 85, 044040 (2012).
\bibitem{cr15} R. Banerjee, D. Roychowdhury, Phys. Rev. D 85, 104043 (2012) arXiv:1203.0118 [gr-qc].
\bibitem{th18}D. Birmingham, Class. Quantum Grav. 16(1999) 1197-1205.
\bibitem{th19} R. G. Cai, A. Wang, Phys. Rev. D 70, 064013, 2004.
\bibitem{th20} R. G. Cai, S. P. Kim, B. Wang, Phys. Rev. D 76, 024011 (2007).
\bibitem{th21}R. G. Cai, Phys. Lett. B 582, 237 (2004). 
\bibitem{horava}P. Horava, Phys. Rev. D 79, 084008 (2009).
\bibitem{horava1} P. Horava, JHEP 03 (2009) 020.
\bibitem{horava2} P. Horava, Phys. Rev. Lett. 102, 161301 (2009).
\bibitem{horava3} M. Visser, Phys. Rev.D 80, 025011 (2009).
\bibitem{Calcagni:2009ar} 
  G.~Calcagni,
  %``Cosmology of the Lifshitz universe,''
  JHEP {\bf 0909}, 112 (2009)
  [arXiv:0904.0829 [hep-th]].\\
%\bibitem{Kiritsis:2009sh} 
  E.~Kiritsis and G.~Kofinas,
  %``Horava-Lifshitz Cosmology,''
  Nucl.\ Phys.\ B {\bf 821}, 467 (2009)
  [arXiv:0904.1334 [hep-th]].\\
  %%CITATION = ARXIV:0904.1334;%%
%\bibitem{Wang:2009rw} 
  A.~Wang and Y.~Wu,
  %``Thermodynamics and classification of cosmological models in the Horava-Lifshitz theory of gravity,''
  JCAP {\bf 0907}, 012 (2009)
  [arXiv:0905.4117 [hep-th]].
  %%CITATION = ARXIV:0905.4117;%%
\bibitem{Lu:2009em} 
  H.~Lu, J.~Mei and C.~N.~Pope,
  %``Solutions to Horava Gravity,''
  Phys.\ Rev.\ Lett.\  {\bf 103}, 091301 (2009)
  [arXiv:0904.1595 [hep-th]].
  %%CITATION = ARXIV:0904.1595;%%
\bibitem{hl3} R. G. Cai, L. M. Cao, N. Ohta, Phys. Rev. D 80, 024003, (2009), arXiv:0904.3670 [hep-th].
\bibitem{Tang:2009bu} 
  J.~-Z.~Tang and B.~Chen,
  %``Static Spherically Symmetric Solutions to modified Horava-Lifshitz Gravity with Projectability Condition,''
  Phys.\ Rev.\ D {\bf 81}, 043515 (2010)
  [arXiv:0909.4127 [hep-th]].\\
  %%CITATION = ARXIV:0909.4127;%%
%\bibitem{Colgain:2009fe} 
  E.~O Colgain and H.~Yavartanoo,
  %``Dyonic solution of Horava-Lifshitz Gravity,''
  JHEP {\bf 0908}, 021 (2009)
  [arXiv:0904.4357 [hep-th]].
  %%CITATION = ARXIV:0904.4357;%%
\bibitem{hl1} Y. S. Myung, Phys. Lett. B 678 (2009) 127–130, arXiv:0905.0957 [hep-th].\\
%\bibitem{Myung:2009dc} 
  Y.~S.~Myung and Y.~-W.~Kim,
  %``Thermodynamics of Horava-Lifshitz black holes,''
  Eur.\ Phys.\ J.\ C {\bf 68}, 265 (2010)
  [arXiv:0905.0179 [hep-th]].
  %%CITATION = ARXIV:0905.0179;%%
\bibitem{hl2} Rong-Gen Cai, Li-Ming Cao, Nobuyoshi Ohta, Phys. Lett. B 679 (2009) 504–509, arXiv:0905.0751 [hep-th].\\
%\bibitem{Majhi:2009xh} 
  B.~R.~Majhi,
  %``Hawking radiation and black hole spectroscopy in Horava-Lifshitz gravity,''
  Phys.\ Lett.\ B {\bf 686}, 49 (2010)
  [arXiv:0911.3239 [hep-th]].
  %%CITATION = ARXIV:0911.3239;%%
\bibitem{stanley1} A. Hankey, H. E. Stanley , Phys. Rev. B 6, 3515 (1972). 
\bibitem{stanley2} H. E. Stanley, Introduction to phase transitions and critical phenomena, Oxford University Press, New York  (1987).
\end{thebibliography}
\end{document}